\documentclass[10pt, aps, pra, 
               amsmath, amssymb, twocolumn,
               preprintnumbers, showpacs,
               raggedbottom,
               floatfix]{revtex4-1}

\usepackage[T1]{fontenc}
\usepackage{textcomp}

\usepackage{etoolbox}
\usepackage{etextools}
\usepackage[acronym]{glossaries}
\glsdisablehyper
\makeglossaries

\usepackage[utf8]{inputenc}

\usepackage[breaklinks]{hyperref}
\usepackage{graphicx}
\graphicspath{{./figures/}}
\usepackage{dcolumn}
\usepackage{calc}

\usepackage[tracking]{microtype}
\SetTracking{ encoding = *, shape = sc }{ 25 }
\usepackage{xcolor}

\AtBeginDocument{\usepackage{bm}}

\usepackage[nice]{nicefrac}

\usepackage{booktabs}

\usepackage{todonotes}
\usepackage{pdfcomment}
\usepackage{color}
\makeatletter
\renewcommand{\todo}[1]{%
  \ifinfloat{\pdfcomment[hoffset=-5pt,icon=Help]{#1}}
            {\@todo{\pdfcomment[hoffset=-5pt,icon=Help]{#1}}}}
\makeatother

\usepackage{siunitx}
\usepackage[version=3]{mhchem}

\newcommand\accentedsymbol[2]{\def#1{#2}}
\accentedsymbol{\vk}{\vect{k}}
\accentedsymbol{\vl}{\vect{l}}
\accentedsymbol{\vp}{\vect{p}}
\accentedsymbol{\vq}{\vect{q}}
\accentedsymbol{\vx}{\vect{x}}
\accentedsymbol{\vy}{\vect{y}}
\let\vr\relax
\accentedsymbol{\vr}{\vect{r}}
\accentedsymbol{\vj}{\vect{\jmath}}
\accentedsymbol{\vv}{\vect{v}}
\accentedsymbol{\vgamma}{\vect{\gamma}}
\accentedsymbol{\valpha}{\vect{\alpha}}
\accentedsymbol{\vbeta}{\vect{\beta}}
\accentedsymbol{\vnabla}{\vect{\nabla}}
\accentedsymbol{\vF}{\vect{F}}
\accentedsymbol{\vR}{\vect{R}}
\accentedsymbol{\oa}{\op{a}}
\accentedsymbol{\ob}{\op{b}}
\accentedsymbol{\oH}{\op{H}}
\accentedsymbol{\oK}{\op{K}}
\accentedsymbol{\oV}{\op{V}}
\accentedsymbol{\otV}{\op{\tilde{V}}}
\accentedsymbol{\psisq}{\abs{\Psi}^2}
\accentedsymbol{\mA}{\mat{A}}
\accentedsymbol{\mB}{\mat{B}}
\accentedsymbol{\mC}{\mat{C}}
\accentedsymbol{\mD}{\mat{D}}
\accentedsymbol{\mE}{\mat{E}}
\accentedsymbol{\mF}{\mat{F}}
\accentedsymbol{\mH}{\mat{H}}
\accentedsymbol{\mM}{\mat{M}}
\accentedsymbol{\mQ}{\mat{Q}}
\accentedsymbol{\mR}{\mat{R}}
\accentedsymbol{\mV}{\mat{V}}
\accentedsymbol{\mX}{\mat{X}}
\accentedsymbol{\mY}{\mat{Y}}
\accentedsymbol{\mDelta}{\mat{\Delta}}
\accentedsymbol{\mchi}{\mat{\chi}}
\accentedsymbol{\mlambda}{\mat{\lambda}}
\accentedsymbol{\metric}{\delta}
\accentedsymbol{\mmetric}{\mat{\delta}}
\accentedsymbol{\ml}{\mat{l}}
\accentedsymbol{\mnu}{\mat{\nu}}

\makeatletter
\newcommand\na[3][\@empty]{%
  {%
    \ifx#1\@empty
      \lowercase{\def\short{\protect\mysc{#2}}}
    \else
      \def\short{#1}
    \fi
    \expandnext{\newacronym{#2}}{\short}{#3}
  }
}
\let\mysc\textsc
\makeatother
\na{GPE}{Gross-Pitaevskii Equation}
\na{QMC}{quantum Monte Carlo}
\na{QHD}{quantum hydrodynamics}
\na{CM}{center-of-mass}
\na{TF}{Thomas-Fermi}
\na{FNQMC}{fixed-node \gls{QMC}}
\na{AFQMC}{auxiliary field \gls{QMC}}
\na{DFT}{density functional theory}
\na{TDDFT}{time-dependent \gls{DFT}}
\na{DMFT}{density-matrix functional theory}
\na{KS}{Kohn-Sham}
\na{LDA}{local density approximation}
\na{BCS}{Bardeen-Cooper-Schrieffer}
\na{BEC}{Bose-Einstein condensate}
\na[\textsc{b}d\textsc{g}]{BdG}{Bogoliubov-de Gennes}
\na{SLDA}{superfluid local density approximation}
\na{EFT}{effective field theory}
\na{LO}{leading order}
\na{NLO}{next-to-leading order}
\na{NNLO}{next-to-next-to-leading order}
\na{ETF}{extended Thomas-Fermi}
\na{DME}{density-matrix expansion}
\na{FFT}{fast Fourier transform}
\na{FFTW}{fastest Fourier transform in the west}
\na{UFG}{unitary Fermi gas}
\na{UGPE}{unitary \gls{GPE}}
\na{UNEDF}{universal nuclear energy density functional}
\na[\textsc{s}\textup{ci}\textsc{dac}]
   {SciDAC}{scientific discovery through advanced
  computing}
\na[\textsc{d}\textup{o}\textsc{e}]{DoE}{Department of Energy}
\na{LDRD}{}
\na{LANL}{Los Alamos National Laboratory}
\na{NERSC}{National Energy Research Science Computing}
\na{EMMI}{ExtreMe Matter Institute}
\na{DFG}{}
\na{GSI}{}
\providecommand{\MMFGRANT}{\textsc{de-fg02-00er41132}}

\providecommand\exclude[1]{}

\providecommand{\abs}[1]{\lvert{#1}\rvert}
\providecommand{\vect}[1]{\vec{#1}}
\DeclareRobustCommand{\order}{\ensuremath{\mathcal{O}}}
\newcommand{\I}{\mathrm{i}}

\usepackage{lettrine}

\usepackage[sc,osf]{mathpazo}
\usepackage[euler-digits,small]{eulervm}
\AtBeginDocument{\renewcommand{\hbar}{\hslash}}

\LettrineOptionsFor{T}{
  lines=3,
  loversize=0.1,
  lraise=-0.03,
  lhang=0.49,
  findent=0.4em,
  nindent=-0.0\LettrineWidth-0.5em}

\LettrineOptionsFor{U}{
  lines=3,
  loversize=0.05,
  lraise=0.03,
  lhang=0.11,
  findent=-0.1em,
  nindent=0.2em}

\LettrineOptionsFor{U}{
  lines=3,
  loversize=0.09,
  lraise=0.01,
  lhang=0.13,
  findent=-0.4em,
  nindent=0.4em,
  slope=0.5em}

\renewcommand{\figurename}{Figure}
\renewcommand{\tablename}{Table}
\makeatletter
\renewcommand{\fnum@figure}{\textbf{\figurename~\thefigure}}
\renewcommand{\fnum@table}{\textbf{\tablename~\thetable}}
\makeatother

\begin{document}

\title{The Unitary Fermi Gas in a Harmonic Trap and its Static Response}

\author{Michael McNeil Forbes}

\affiliation{Institute for Nuclear Theory, University of Washington,
  Seattle, Washington 98195--1550 USA}
\affiliation{Department of Physics, University of Washington, Seattle,
  Washington 98195--1560 USA}

\date{\today}

\newcommand{\cSR}{c_{\chi}}
\newcommand{\dlambda}{\,\delta\lambda}
\begin{abstract}
  \glsresetall
  \noindent
  We use harmonically trapped systems to find the leading gradient corrections
  of the \gls{SLDA} -- a \gls{DFT} describing the \gls{UFG}.  We find the
  leading order correction to be negative, and predict the $q^2$ coefficient of
  the long-range static response $\cSR = \num{1.5(3)}$ -- a factor of two
  smaller than predicted by mean-field theory -- thereby establishing a new
  and experimentally measurably universal constant.
\end{abstract}
\preprint{\textsc{int-pub-12-057}}

\maketitle
\glsresetall

\lettrine{U}{niversally} describing two-component Fermi systems with short-range
interactions of infinite scattering length $a_s \rightarrow \infty$, the
\gls{UFG}~\cite{Zwerger:2011} not only approximates the dilute neutron matter
found in neutron stars~\footnote{Dilute neutron matter is also well modelled by
  the \gls{UFG}~\cite{Gezerlis;Carlson:2008-03} as a consequence of the
  unnaturally large neutron-neutron scattering length:
  $a_{nn}\approx\SI{-18.9(4)}{fm}$~\cite{Gonzalez-Trotter:2006, *Chen:2008}
  while densities are on the order of $k_F^{-1} \sim \SI{1}{fm}$. The effective
  s range, however, is not small: $r_{nn}\approx
  \SI{2.75(11)}{fm}$~\cite{Miller:1990} implying $k_Fr_e \approx 3$.  Thus,
  range corrections must generally be included to quantitatively describe these
  systems.  Despite this complications, the qualitative properties of dilute
  neutron matter are well described by the \gls{UFG}.}, but is directly realized
in cold-atom systems~\footnote{The scattering length of dilutely trapped alkali
  atoms may be tuned $\abs{a_s} \gg k_F^{-1} \gg r_e$ such that the \gls{UFG}
  may be directly realized in systems of cold atoms (see Refs.~\cite{IKS:2008,
    *giorgini-2007, *Zwerger:2011} for reviews).}, allowing experiments to
benchmark many-body techniques used to study astrophysical
phenomenology. Despite the simplicity of the system -- the lack of
length-scales, for example, implies that the equation of state $\mathcal{E}(n_+)
\propto \smash{n_+^{5/3}}$ -- the system is strongly interacting and admits no
known perturbative expansions: A quantitative description requires experiments
or \textit{ab initio} computations.

\textit{Ab initio} techniques, however, can only address a few questions --
direct \gls{QMC} simulations, for example, can study systems with at most a few
hundred particles.  It is therefore imperative to benchmark computationally
tractable models of macroscopic phenomena so that they can be used to answer
outstanding phenomenological questions, such as the origin of glitching in
neutron stars~\cite{Anderson:1975}.

\Gls{DFT} is an in principle exact approach, widely used in nuclear physics
(see~\cite{Drut:2010kx} for a review), and in quantum chemistry to describe
normal (i.e., non-superfluid) systems.  It provides a framework capable of
assimilating \textit{ab initio} and experimental results into a computationally
tractable and predictive framework.  In this letter, we extend one such
\gls{DFT} -- the \gls{SLDA} -- to describe the inhomogeneous behaviour of
harmonically trapped systems.  We use \gls{DFT} to analyse recent experimental
and theoretical results, noting discrepancies and the asymptotic behaviour
toward the thermodynamic limit, and establish the leading order gradient
corrections to the \gls{SLDA} which we find to be negative.  The \gls{SLDA} then
uniquely predicts the low-energy static response of the \gls{UFG} to quadratic
order, which is crucial for a proper low-energy description of the \gls{UFG}:
this therefore makes significant progress towards a predictive framework for
studying superfluid phenomenology.

At low-energies, the \gls{UFG} can be characterized by a superfluid \gls{EFT}
describing phonon dynamics~\cite{SW:2006}.  The \gls{EFT} admits a controlled
power-counting scheme: The \gls{LO} contains a single dimensionless parameter --
the Bertsch parameter~\cite{mbx, *Baker:1999:PhysRevC.60.054311,
  *baker00:_mbx_chall_compet} $\xi$ which characterizes the equation of state
$\mathcal{E}(n) = \xi \mathcal{E}_{FG}(n_+)$ where $\mathcal{E}_{FG} = 3/5 n_{+}
E_{F}$ is the energy density of a free Fermi gas with the same total density
$n_{+} = k_F^3/(3\pi^2)$, and $E_{F} = \hbar^2 k_F^2/2m$ is the Fermi energy.
At \gls{NLO}, two additional dimensionless coefficients~\cite{SW:2006,
  Manes:2009} appear which characterize the static and dynamic low-frequency and
low-momentum response.  We shall address only the value of the
static response $\chi_q$ as defined by adding small external potential $\delta
V_q(x) = \delta\cos(qx)$ to a homogeneous system:
\begin{gather}
  \delta n_+(x) = \chi_q\delta V_q(x) + \order(\delta^2).
\end{gather}
To \gls{NLO} in the superfluid \gls{EFT}, the response is
\begin{gather}
  \label{eq:SonWingateStaticResponse}
  \chi(q) = \frac{-mk_F}{\hbar^2\pi^2\xi}\left[
    1 - \frac{\cSR}{12\xi}\frac{q^2}{k_F^2} + \order(q^4\ln{q})
  \right]
\end{gather}
where $\cSR$ is a universal dimensionless constant \footnote{In the notations
  of~\cite{SW:2006}, $\cSR \equiv -6\pi^2(2\xi)^{3/2}(2c_1 - 9c_2)$ while in the
  notations of~\cite{Manes:2009}, $\cSR \equiv -12\pi^2(2\xi)^{3/2}c_1$.}.  This
normalization for $\cSR$ is numerically close to unity ($\cSR=1$ for
non-interacting fermions), appears simply in the energy of trapped fermions (see
Eq.~\eqref{eq:Etrap_x}), and is independent of the $\xi$ and pairing parameters
in the \gls{SLDA}. The other universal constant $c_\omega$ describes low-energy
dynamical properties, and enters through the phonon dispersion relation
\footnote{In~\cite{SW:2006}, $c_\omega \equiv -6\pi^2(2\xi)^{3/2}(2c_1 + 3c_2)$
  while in~\cite{Manes:2009}, $c_\omega \equiv -12\pi^2(2\xi)^{3/2}(c_1 -
  3c_2)$.}  $\omega_q/(qc_s) = 1 + c_\omega q^2/(24\xi k_F^2) +
\order(q^4\ln{q})$ where $c_s = \hbar k_F\sqrt{\xi/3}/m$ is the speed of sound.

While many techniques have been employed to calculate the Bertsch parameter $\xi
\approx 0.37$ (see~\cite{Endres:2012} for a survey), \gls{NLO} coefficients have
only been considered in a few cases: The
$\epsilon$-expansion~\cite{Rupak:2008fk} (expanding in spatial dimension:
$\epsilon = 4 - d$) gives $\cSR \approx 8/5 + \order(\epsilon^2)$ and $\cSR
\approx c_\omega + \order(\epsilon^2)$ \footnote{To compare
  with~\cite{Rupak:2008fk}, their $c_s = c/2\xi$.}, while \gls{BdG} mean-field
theory~\cite{Manes:2009} finds $\cSR = 7/3$ and $c_\omega = 0.7539$.

The \gls{EFT} breaks down for small systems and near the boundary of clouds, so
to connect with finite-size \gls{QMC} calculations, we use \gls{DFT}.  The
Hohenberg-Kohn theorem~\cite{HK:1964} asserts the existence of a universal
functional of the density alone whose minimum describes the ground state of the
\gls{UFG}. The exact form of this functional is non-local and unknown, but a
local formulation -- an \gls{ETF} functional~\cite{Rupak:2008fk, Salasnich:2008,
  *Salasnich:2008E, *Manzoni:2010, Salasnich:2010, Ancilotto:2012,
  *Ancilotto:2012a, Salasnich:2012a} -- describes well some energetic and
dynamical aspects of the \gls{UFG}.  It fails, however, to properly describe
finite-size effects in homogeneous systems~\cite{FGG:2010, Forbes:2012}.

To describe these properties we use the Kohn-Sham formulation~\cite{Kohn:1965fk}
which includes an auxiliary kinetic density $\tau_+$.  While this is
formally equivalent to the Hohenberg-Kohn formulation, the addition of a
kinetic density allows a local formulation to describe finite-size features of
the system.  In particular, the finite-size properties of non-interacting
systems are exactly reproduced.  Interacting versions have been
considered~\cite{Papenbrock:2005fk, Bhattacharyya:2006}, but one finds that the
finite-size effects are not properly suppressed~\cite{FGG:2010, Forbes:2012}.
The suppression can be realized by include an additional auxiliary anomalous
density, $\nu$, representing the pairing field~\cite{Bulgac:2007a, *Bulgac:2011,
  FGG:2010, Forbes:2012}, resulting in the \gls{SLDA}:
\begin{gather}
  \mathcal{E}_{\textsc{slda}} =
  \frac{\hbar^2}{m}\left(
    \frac{\alpha}{2}\tau_{+} +  g\nu^{\dagger}\nu\right) 
  + \beta \mathcal{E}_{FG}(n_+)
  + \frac{\hbar^2\dlambda}{8m}\frac{(\vnabla n_+)^2}{n_+},\nonumber\\
  g^{-1} = \frac{n_{+}^{1/3}}{\gamma} - \frac{k_c}{2\pi\alpha}
  \label{eq:DF_SLDA}
\end{gather}
Here $\alpha$ is the inverse effective mass, $\beta$ is the self-energy,
$\gamma$ controls the pairing, and $\dlambda$ characterizes the leading order
gradient term (known as a Weizsäcker correction).  The unitary limit is realized
when we take the wave-vector cutoff $k_c\rightarrow\infty$ to infinity (see
Ref.~\cite{Bulgac:2011} for details). In homogeneous systems, the gradient
corrections vanish, and one can use the equations in the thermodynamic limit to
replace the parameters $\beta$ and $\gamma$ by the more physically relevant
quantities $\xi_S$ and $\eta = \Delta/E_F$, where $\Delta$ is the pairing gap
(see the appendix of~\cite{Forbes:2012} for details).  When applied to
inhomogeneous systems, however, one must hold the parameters $\beta$ and
$\gamma$ fixed to define the functional.

This form~\eqref{eq:DF_SLDA} subsumes earlier \glspl{DFT}. In particular, the
well-studied \gls{BdG} mean-field equations are reproduced with unit effective
mass $\alpha=1$, $\gamma^{-1}=0$, no Hartree term $\beta=0$, and no gradient
corrections $\dlambda=0$. The Kohn-Sham form discussed
in~\cite{Papenbrock:2005fk, Bhattacharyya:2006} neglects the $\nu=0$, while the
\gls{ETF} form~\cite{Rupak:2008fk, Salasnich:2008, *Salasnich:2008E,
  *Manzoni:2010} is reproduced if one neglects both the anomalous density $\nu$
and the kinetic density $\tau_+$.  As discussed in~\cite{FGG:2010}, none of
these restricted forms can even qualitatively characterize the finite-size
effects, but we still consider the \gls{ETF} functional as it is much easier to
solve numerically while retaining the asymptotic properties of trapped systems:
\begin{gather}\label{eq:ETF}
  \mathcal{E}_{\acrshort{ETF}} =
  \xi\mathcal{E}_{FG}(n_+) 
  + \frac{\hbar^2(1/4 + \dlambda)}{8 m}\frac{(\vnabla n_+)^2}{n_+}.
\end{gather}
The leading gradient term here derives from a semi-classical expansion of the
kinetic energy~\cite{Brack:1997,*Ring:2004} with an additional Weizsäcker
correction $\dlambda$.  Superfluid hydrodynamic
phenomenology~\cite{Ancilotto:2012, *Ancilotto:2012a} and vortex dynamics
(appendix~\ref{sec:hydr-with-etf}) suggest that $\dlambda = 0$.  The
resulting \gls{ETF} is completely determined by the value of $\xi$.  A simple
calculation~\cite{Salasnich:2012a} shows that the \gls{ETF} model has $\cSR =
c_\omega = 9/4 + 9\dlambda = 9/4$.

The \gls{SLDA} was originally constrained by \gls{QMC} calculations of the
continuum state, and validated with \gls{QMC} calculations in a harmonic
trap~\cite{Chang;Bertsch:2007-08, Blume;Stecher;Greene:2007-12}.  These
validations, however, provided only a weak test of the \gls{SLDA} form.  In
particular, the symmetric thermodynamic limit does not provide enough
information to constrain the effective mass, and the original variational trap
results were not sufficiently accurate to exhibit the appropriate scaling in the
thermodynamic limit~\cite{FGG:2010, Forbes:2012}.

Recently, experimental and \textit{ab initio} \gls{QMC} results for homogeneous
matter in the continuum and in periodic boxes were used to more rigorously test
the form of the \gls{SLDA}~\cite{Forbes:2012}: the best fit to current unbiased
results is consistent with $\xi = \num{0.3742(5)}$, $\alpha=1.104(8)$, and
$\eta=0.651(9)$.  Here we estimate the leading order gradient correction
$\dlambda$ by reconsidering the energies of trapped systems.

The static response in the thermodynamic limit can be calculated using the same
techniques as in the \gls{BdG}~\cite{Manes:2009} and one
finds~\cite{Forbes:2012d}
\begin{gather}
  \cSR = \tfrac{7}{3}\alpha + 9\dlambda,
\end{gather}
which is independent of $\xi$ and $\eta$. This demonstrates how the effective
mass and gradient corrections play a similar role, as pointed out
in~\cite{Bulgac:1995, Bhattacharyya:2005}.

From the \gls{NLO} superfluid \gls{EFT}~\cite{SW:2006, Manes:2009}, one finds
the energy of the \gls{UFG} in an isotropic harmonic trap with trapping
frequency $\omega$ to depend on the coefficients $\xi$ and
$\cSR$:
\begin{gather*}
  \frac{E}{\hbar\omega} = \frac{\sqrt{\xi}}{4}(3N_+)^{4/3}\left[
    1 - \frac{\cSR}{2\xi(3N_+)^{2/3}} + \order\left(\frac{1}{N_+^{7/9}}\right)
  \right].
\end{gather*}
This form naturally suggests the abscissa $x=(3N_+)^{-2/3}$ so that the
asymptotic behavior of $E$ is linear: we prefer to use the square of the energy
$E^2$,
\begin{gather}
  \label{eq:Etrap_x}
  y = \frac{16E^2}{\hbar^2\omega^2(3N_+)^{8/3}} = \xi + \cSR x 
  + \order\left(x^{7/6}\right)),
\end{gather}
as $\xi$ appears on the intercept, and $\cSR$ appears directly.  It is
interesting that, in the non-interacting system, shell-effects appear at the
same linear order $x$, leading to a fundamental uncertainty in the coefficient
$\frac{2}{3} \leq \cSR \lessapprox 1.7$. Pairing suppresses these shell effects,
yielding a well-defined asymptotic slope $\cSR$, which can also be determined
from the semi-classical approximation (see appendix~\ref{sec:semicl-expas}).

In Fig.~\ref{fig:QMC} we display \gls{QMC} results for trapped unitary
systems. The dotted lines guide the eye through several variational bounds
obtained using \gls{FNQMC} calculations.  In these methods, one avoids any sign
problem by sampling a restricted set of wavefunctions with the same nodal
structure as an initial reference ansatz.
\exclude{If the nodal structure
matches that of the ground state, then these methods are exact, but in general,
they only provide an upper bound.}
By improving the ansatz and varying the parameters, these bounds have come down
over time, and the lowest (green) curve~\cite{Forbes:2012} represents the best
bound to date.  Note that this is the only set of results that demonstrates the
expected linear scaling~\eqref{eq:Etrap_x} predicted by the effective theory.
We suspect that numerical issues or the nature of the ansatz in the other cases
introduced spurious lengths scales that violate this scaling
(see~\cite{Forbes:2012} for further discussion.)

\begin{figure}[tb]
  \includegraphics[width=\columnwidth]{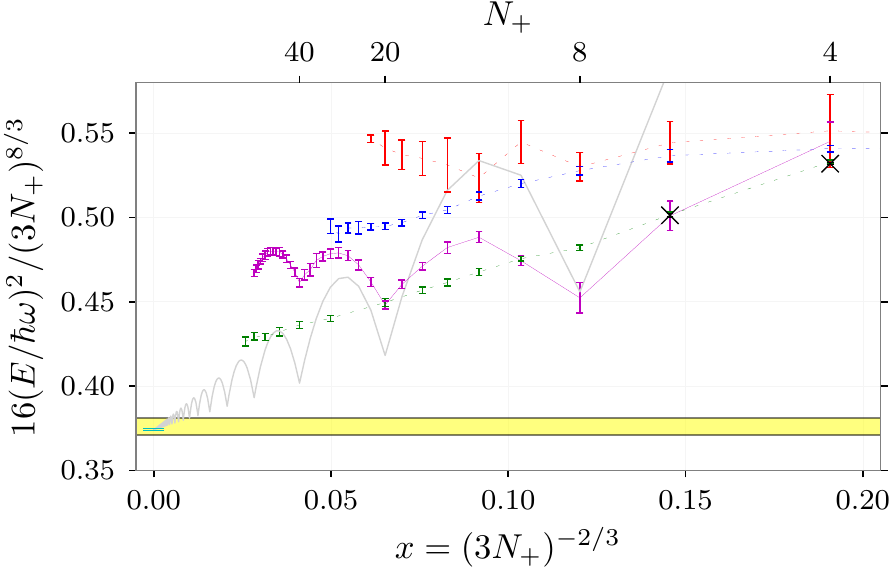}
  \caption{(color online) 
    Various \gls{QMC} and experimental results for trapped systems (with lines
    to guide the eye).  The results with dotted lines are from \gls{FNQMC}
    calculations that provide only upper bounds due to the nodal constraint.
    From top to bottom these results are from~\cite{Chang;Bertsch:2007-08}
    (blue), \cite{Blume;Stecher;Greene:2007-12} (red), and~\cite{Forbes:2012}
    (green).  The (magenta) points with solid lines are from a lattice
    calculation~\cite{Endres:2011} that is in principle unbiased.  The large
    (black) crosses for $N_+ \in \{4, 6\}$ are from~\cite{Blume:2010}.  The
    solid light (grey) curve shows the shell structure for free fermions in the
    trap (the curve has been shifted down from $1$ to facilitate
    comparison). Finally, we include the latest experimental value for
    $\xi=\num{0.376(5)}$ from~\cite{Ku:2011} as a (yellow) band and the best fit
    value of $\xi=\num{0.3742(5)}$ to all homogeneous \textit{ab initio}
    \gls{QMC} results from~\cite{Forbes:2012} at $x=0$.  Coordinates have been
    scaled as in~\eqref{eq:Etrap_x} to demonstrate the scaling.  The
    corresponding particle numbers $N_+$ are listed along the top axis, and
    emphasize the closed shells which occur for $N_+ \in \{2, 8, 20, 40, 70\}$.
    (Note that all methods agree for the point $N_+=2$ (not shown) which admits
    an analytic solution.)}
  \label{fig:QMC} 
\end{figure}

The solid (magenta) line guides the eye through calculations based on lattice
techniques~\cite{Endres:2011}.  In principle, these are unbiased \textit{ab
  initio} results, but it is somewhat troubling that most lie significantly
above the variational bounds.  They also display large shell effects that are
virtually absent in the \gls{FNQMC} results. For comparison, we have included
the energies of free particles shown in Fig.~\ref{fig:free} as a light (grey)
curve, shifted down from $1$ to facilitate comparison. As we shall see, although
the non-interacting \gls{SLDA} reproduces these shell-effects, the interacting
\gls{SLDA} exhibits a marked lack of shell effects, consistent with the
\gls{FNQMC}. The lattice results thus seem qualitatively inconsistent with the
others.

A third variational technique~\cite{Blume:2010} based on a correlated Gaussian
approach provides very tight bounds, but is limited to small systems.  These are
shown as (black) crosses for $N_+ \in \{2, 4, 6\}$. Unfortunately, at these
three points, all methods agree, and significant discrepancies between the
lattice and \gls{FNQMC} results only appear at larger $N_+$.

Finally, we include the latest experimental results as a
(yellow) band~\cite{Ku:2011} and the best fit value of $\xi = \num{0.3742(5)}$
to all homogeneous \textit{ab initio} \gls{QMC} results from~\cite{Forbes:2012}
at $x=0$.  According to the scaling~\eqref{eq:Etrap_x}, the results should
approach this point.

\begin{figure}[tb]
  \includegraphics[width=\columnwidth]{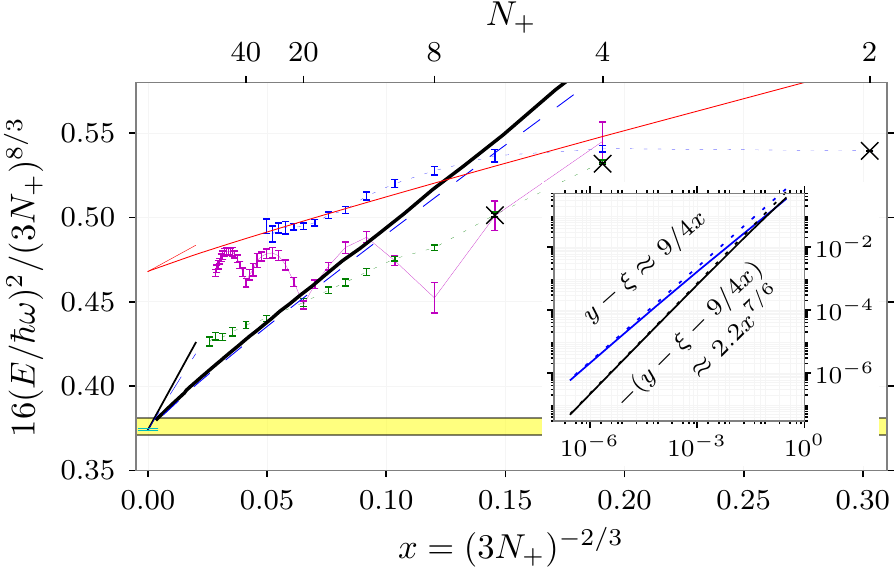}
  \caption{(color online) 
    \Gls{DFT} models compared with data from Fig.~\ref{fig:QMC}. The thick solid
    (black) curve is the \Gls{SLDA} with parameters $\xi = \num{0.3742(5)}$,
    $\alpha=1.104(8)$, and $\eta=0.651(9)$ but no gradient corrections
    $\dlambda=0$.  The thin dashed (blue) curve is the \gls{ETF} with the same
    $\xi = 0.3724$ and $\dlambda=0$.  For comparison, the upper thin solid
    (red) curve corresponds to the best fit \gls{ETF} model ($\xi=0.468$,
    $\dlambda = -0.164$) described in~\cite{Salasnich:2008,
      *Salasnich:2008E, *Manzoni:2010} to the \gls{FNQMC}
    result~\cite{Blume;Stecher;Greene:2007-12}.  The linear asymptotic forms $y
    = \xi + \cSR x$ with $\cSR = 7\alpha/3 \approx \num{2.58}$ (\gls{SLDA}) and
    $\cSR = \num{2.25}$ (\gls{ETF}) are shown as short thin lines extending from
    $x=0$. The inset is a log-log plot of the deviation the \gls{ETF} model
    makes from the asymptotic form $y = \xi + 9/4 x + a
    x^{7/6} + \cdots$ where $ a\approx -2.2$: the solid curves are the
    deviations ($y-\xi$ above (blue) and $-(y - \xi - 9/4x)$
    below (black)) while the dotted lines are the expected order
    ($9/4 x$ above and $ax^{7/6}$ below).}
  \label{fig:ETF_SLDA} 
\end{figure}

In figure~\ref{fig:ETF_SLDA} we overlay the \gls{DFT} results for the trapped
systems.  The best fit \gls{SLDA} without gradient corrections (solid) and the
\gls{ETF} with the same $\xi=0.3742$ and $\dlambda = 0$ (dashed) have almost
exactly the same structure, demonstrating the ability of the kinetic term in the
\gls{SLDA} to model the gradient effects of the \gls{ETF}.  While the \gls{DFT}
results appear to approach a linear asymptotic form, the actual slope $\cSR$
determined from the static response is only realized for extremely large $N_+$.
This is expected due to the very slight suppression of higher order terms $\sim
x^{7/6}$: to suppress these corrections by an order of magnitude requires $0.1
\approx x^{1/6}$ which implies that $N_+ \gtrapprox 10^8$.  Thus, there is
virtually no hope of directly extracting the value of $\cSR$ from the \gls{QMC}
simulations presented in Fig.~\ref{fig:QMC} without using a model to extrapolate
to large $N_+$.  Both \gls{DFT} models ultimately exhibit this behaviour as we
demonstrate with the \gls{ETF} in the inset.

\begin{figure}[tb]
  \includegraphics[width=\columnwidth]{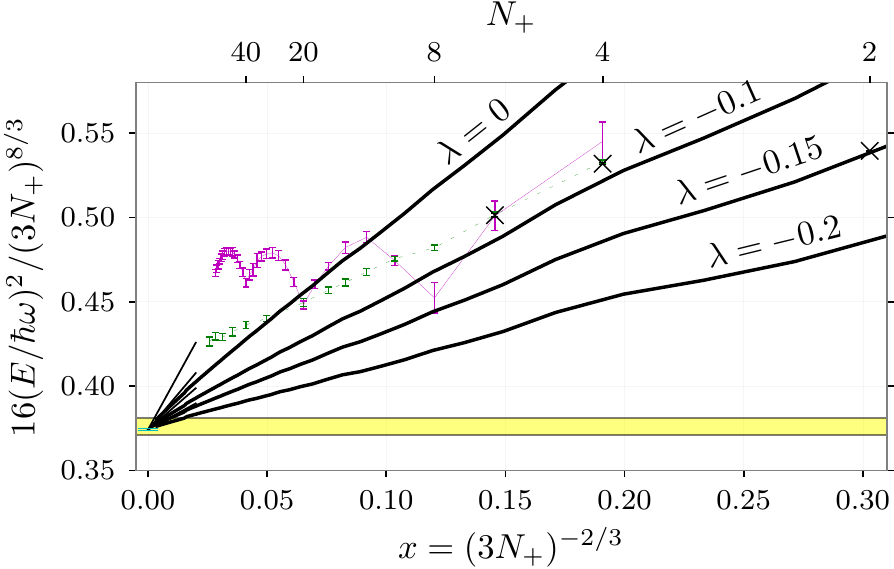}
  \caption{(color online) 
    \Gls{SLDA} model as in Fig.~\ref{fig:ETF_SLDA} with various values of the
    gradient corrections $\lambda$ added perturbatively.  }
  \label{fig:SLDA_grad} 
\end{figure}

Note that the \gls{SLDA} predicts higher energies for systems with small
particle numbers -- especially the $N_+ \in \{2, 4, 6\}$ systems where all
\textit{ab-initio} methods agree.  To correct for this, the leading gradient
correction needs to be negative $\lambda < 0$.  We explore these effects in
Fig.~\ref{fig:SLDA_grad} by perturbatively including the gradient for various
values of $\lambda$.  (Note: these corrections are less than 15\% for all
systems as shown in Fig.~\ref{fig:SLDA_pert}.  The remaining corrections from a
fully self-consistent solution will not significantly alter these results.)  One
could also increase the effective mass, but to match even the modest correction
of $\lambda \lessapprox -0.1$ requires $m_{\text{eff}} \gtrapprox 1.4 m$ which
spoils the description of homogeneous systems~\cite{FGG:2010, Forbes:2012} and
quasi-particle dispersions~\cite{Carlson;Reddy:2008-04}. Higher order gradient
corrections might help, however, there are several different gradient
corrections -- each requiring additional coefficients -- and insufficient
\textit{ab-initio} results to constrain these.  These neglected terms will not
affect the coefficient $\cSR$.  Finally, there is the possibility that the
functional could be generalized as discussed in~\cite{FGG:2010}, but the success
of the three-parameter \gls{SLDA}~\cite{Forbes:2012} suggests that corrections
along this line would be small.

We note that a negative gradient correction is somewhat surprising since a naïve
expansion of an attractive non-local interaction $-V(x-y)n(x)n(y)$ yields a
positive gradient correction (see appendix~\ref{sec:naive-grad-corr}).

Thus, we conclude from Fig.~\ref{fig:SLDA_grad} that the \gls{SLDA} will require
$\lambda \approx -0.12(3)$ to describe both homogeneous boxes and trapped
systems.  The \gls{SLDA} therefore predicts
\begin{gather}
  \label{eq:c_value}
  \cSR \approx \num{1.5(3)}
\end{gather}
where the error is approximate (not to be taken as a standard deviation).  This
is about half the value predicted by the \gls{BdG} mean-field calculation,
though the effect in the static response~\eqref{eq:SonWingateStaticResponse},
$\propto \cSR/\xi$, is cancelled by the excessively large \gls{BdG} value of
$\xi=0.5906$.  Intuitively, $\cSR/\xi$ is four times larger than in the
non-interacting system, indicating that momentum-dependent density fluctuations
are suppressed by interaction, though this effect is mostly due to the reduced
value of $\xi$.

As discussed, $\cSR$ cannot be directly extracted from the \gls{QMC} results
without a model capable of extrapolating well into the thermodynamic limit.  To
extract $\cSR$ more directly, one should consider systems that minimize the
sensitivity to the breakdown of the \gls{EFT} at the boundary of the system.  To
do this, consider how the density $n\equiv n(\vx)$ depends on a smoothly varying
potential $V\equiv V(\vx)$:
\begin{gather}
  n = n_{\acrshort{TF}}(\vx)\left\{1
    - 
    \frac{\cSR}{64}
    \frac{(\vnabla V)^2
      + 4(\mu-V)\nabla^2V
    }{(\mu - V)^3}
    \frac{\hbar^2}{m}
    + \cdots
  \right\},\nonumber\\
  n_{\acrshort{TF}}(\vx) = \frac{8\bigl[\mu - V\bigr]^{3/2}}{3\pi^2(2\xi)^{3/2}}
  \left(\frac{m}{\hbar^2}\right)^{3/2}.
\end{gather}
This is valid to \gls{NLO} in a static background with constant phase (i.e.\@
not near a vortex).  Thus, one can directly search for deviations from the
\gls{TF} profile that are sensitive to gradients in the potential.  Applying a
modulation $V \propto \cos(qx)$ directly probes the static response and is
feasible in \gls{QMC} simulations.  Optical latices could be used similarly in
experiments, however, measuring the density to sufficient accuracy is likely to
be a challenge -- to compensate for the numerical suppression $\cSR/64$ while
avoiding contamination from higher-order terms will require percent level
accuracy.  Experiments should thus probably retain traps with axial or spherical
symmetry so they can benefit from averaging techniques like the inverse Abel
transform to reduce noise in $n(\vx)$.  Adding a small dimple to the core of the
trap with varying widths will allow experiments to probe the response while
retaining the benefits of averaging to reduce noise.

Confirming the value of $c_\chi$ will provide an important validation of the
\gls{SLDA} functional, and provides another benchmark for models of the
\gls{UFG}: To reliably predict low-energy behaviour of the \gls{UFG}, a model
should reproduce the \gls{LO} and \gls{NLO} coefficients -- $\xi$, $c_\chi$, and
$c_\omega$ -- of the superfluid \gls{EFT}

\begin{acknowledgments}
  \noindent
  We thank D.~Blume, A.~Bulgac, R.J.~Furnstahl, S.~Gandolfi, A.~Gezerlis,
  S.~Reddy, R.~Sharma, and D.T.~Son for useful discussions.  This work is
  supported by \textsc{us} \gls{DoE} grant \MMFGRANT.
\end{acknowledgments}

\clearpage
\appendix

\section{Contradictory QMC Results}\noindent
As noted in Fig.~\ref{fig:QMC}, there are unresolved contradictions in the
\gls{QMC} results for larger $N_+$ with the lattice results~\cite{Endres:2011}
exceeding the \gls{FNQMC} variational bounds~\cite{Forbes:2012}.  It would be
useful to have an alternative method calculate the energies for $N_+=8$ and
$N_+=12$ to resolve between these:
\begin{align}
  \frac{E_{\text{lat}}}{\hbar\omega} &= 11.64^{+0.106}_{-0.124}
  & \frac{E_{\acrshort{FNQMC}}}{\hbar\omega} &\leq \num{12.01(2)}
  \tag{$N_+ = 8$}
  \\
  \frac{E_{\text{lat}}}{\hbar\omega} &= 20.765^{+0.045}_{-0.093}
  & \frac{E_{\acrshort{FNQMC}}}{\hbar\omega} &\leq \num{16.07(2)}
  \tag{$N_+ = 12$}
\end{align}
\exclude{ --
$\smash{E^{\text{lat}}_{8} = 11.64^{+0.106}_{-0.124}}\hbar\omega$ and
$\smash{E^{\text{lat}}_{12} = 20.765^{+0.045}_{-0.093}}\hbar\omega$ -- and the
 -- $E\smash{^{\acrshort{FNQMC}}_{8}} = \num{12.01(2)}\hbar\omega$
and $\smash{E^{\acrshort{FNQMC}}_{12}} = \num{16.07(2)}\hbar\omega$.}

\section{Semiclassical Expasion}\label{sec:semicl-expas}\noindent
The superfluid \gls{EFT} is closely related to the well-studied semiclassical
expansion (in $\hbar$)~\cite{Brack:1997,*Ring:2004} for non-interacting systems,
and one can derive similar expressions to those arising from the \gls{EFT}.  The
utility of the \gls{EFT} is to organize the universal coefficients for the
interacting superfluid system.

\begin{figure}[h]
  \includegraphics[width=\columnwidth]{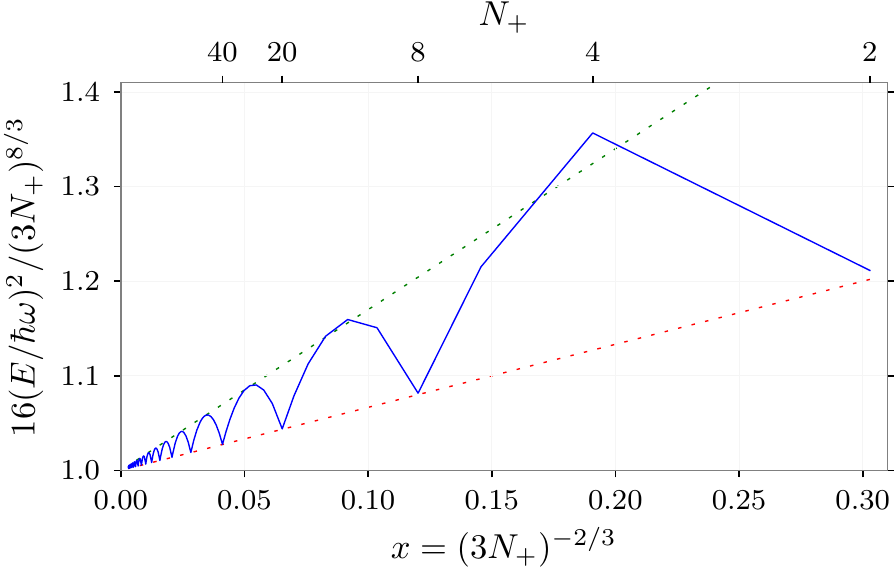}
  \caption{(color online) 
    Shell effects in the trapped non-interacting two-component gas. The scaling
    is the same as in Fig.~\ref{fig:QMC}.  The asymptotic bounds $\frac{2}{3}
    \leq \cSR \lessapprox 1.7$ have been drawn as dotted lines.}
  \label{fig:free} 
\end{figure}
The shell effects for non-interacting systems is shown in Fig.~\ref{fig:free}
(this was shifted down to match $\xi$ for comparison in Fig.~\ref{fig:QMC}).
The semiclassical expansion systematically organizes contributions from volume
effects, surface effects, periodic orbits, etc.  It thus provides some insight
into the breakdown of the \gls{EFT}: shell for the harmonic trap, for example,
effects appear at the same order as the \gls{NLO} corrections.  These are shown
in Fig.~\ref{fig:free} demonstrating a fundamental uncertainty in the slope
$\frac{2}{3} \leq \cSR \lessapprox 1.7$. Considering the static response allows
one to extract the non-interacting value of $\cSR = 1$, which lies in this band.
The fact that these corrections appear at the same order is related to nearby
breakdown of the \gls{EFT} formula, and the subsequent slow approach to the
asymptotic scaling in the thermodynamic limit. What is non-trivial is that
pairing acts to sufficiently suppress these shell effects so that a well defined
slope emerges in harmonically trapped systems. Perhaps this can be explained
within the semiclassical theory, but the author is not aware of such a
discussion in the literature.

\section{Perturbative Gradient Corrections}
\noindent
In Fig.~\ref{fig:SLDA_grad} we show the size of the gradient corrections
$E_{\text{grad}}$ as a percent of the total energy $E$ for the $\dlambda =
-0.15$ required to bring the \gls{SLDA} in rough accordance with the smallest
trapped systems.  This demonstrates that the gradient corrections may be
included perturbatively, simplifying the numerical calculations. 

\begin{figure}[h]
  \includegraphics[width=\columnwidth]{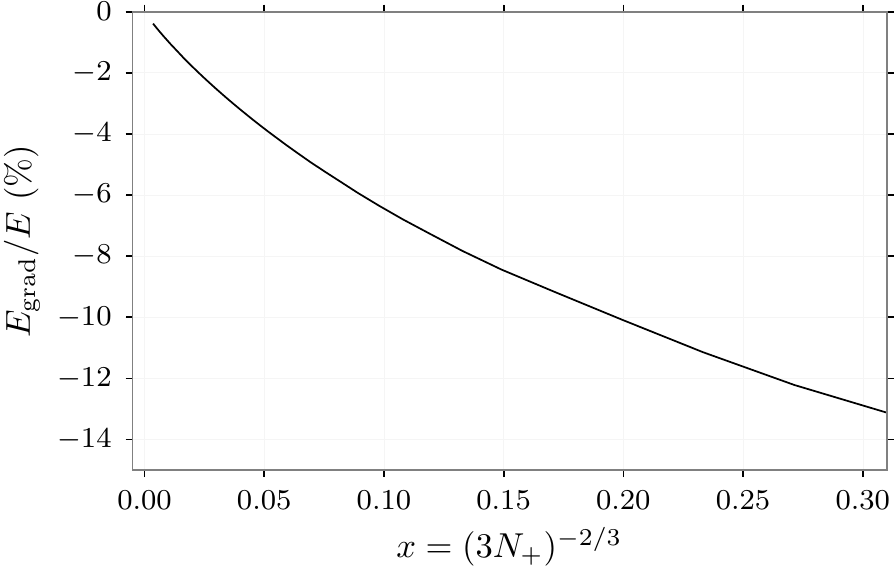}
  \caption{With $E_{\text{grad}}/E$ (in percent) for $\dlambda=-0.15$,
    demonstrating that the correction is indeed perturbative.
  }
  \label{fig:SLDA_pert} 
\end{figure}

In order to perform a fully self-consistent calculation of the gradient
corrections, one must modify the single-particle self-energy to include a term
\begin{gather*}
  U_{\text{grad}} = - \frac{\hbar^2\dlambda}{8m}
  \vnabla\cdot\left(\frac{\vnabla n_+}{n_+}\right)
  = \frac{\hbar^2\dlambda}{8m}
  \left(4\frac{\nabla^2\sqrt{n_+}}{\sqrt{n_+}}
    - \frac{\nabla^2 n_+}{n_+}\right).
\end{gather*}
Accurately computing the derivatives -- especially in the tails of the cloud --
presents a mild numerical challenge, and will only alter the energies at the
percent level, and so is not required for the present analysis.  This
self-consistency will be important, however, if one wants to compare the density
profile of the smallest trapped systems with \gls{QMC} results.

\section{Asymptotic Behaviour}
\begin{figure}[t]
  \includegraphics[width=\columnwidth]{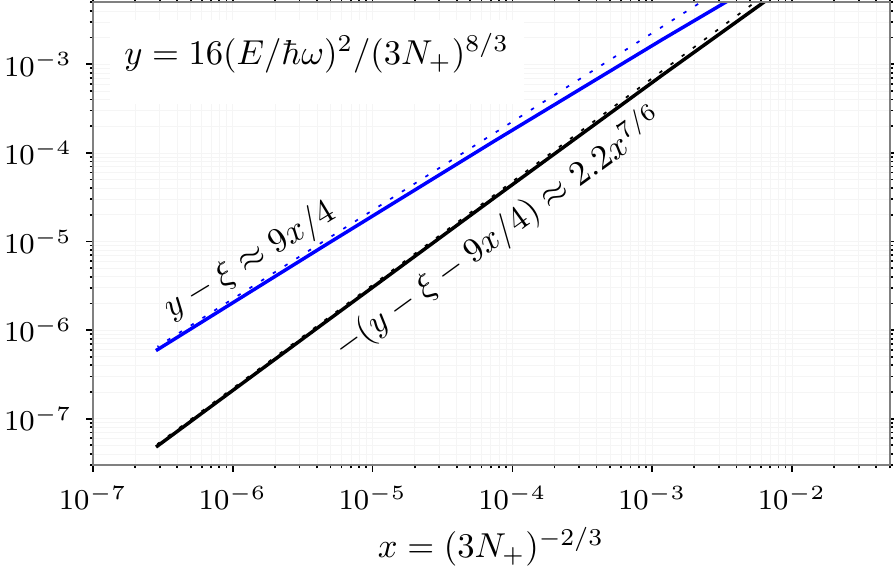}
  \caption{(color online) 
    Asymptotic behaviour of the \gls{ETF} model with $\xi=\num{0.3742}$ and
    $\dlambda=0$: $y = \xi + 9 x/4 + a x^{7/6} + \cdots$ where $a\approx -2.2$.
    We plot the deviations $y-\xi$ (upper solid blue curve) and $-(y - \xi -
    9x/4)$ (lower solid black curve) along with the next terms $9x/4$ (upper
    dotted blue line) and $2.2 x^{7/6}$ (lower dotted black line) to demonstrate
    the asymptotic powers.}
  \label{fig:ETF_asympt} 
\end{figure}

\begin{figure}[t]
  \includegraphics[width=\columnwidth]{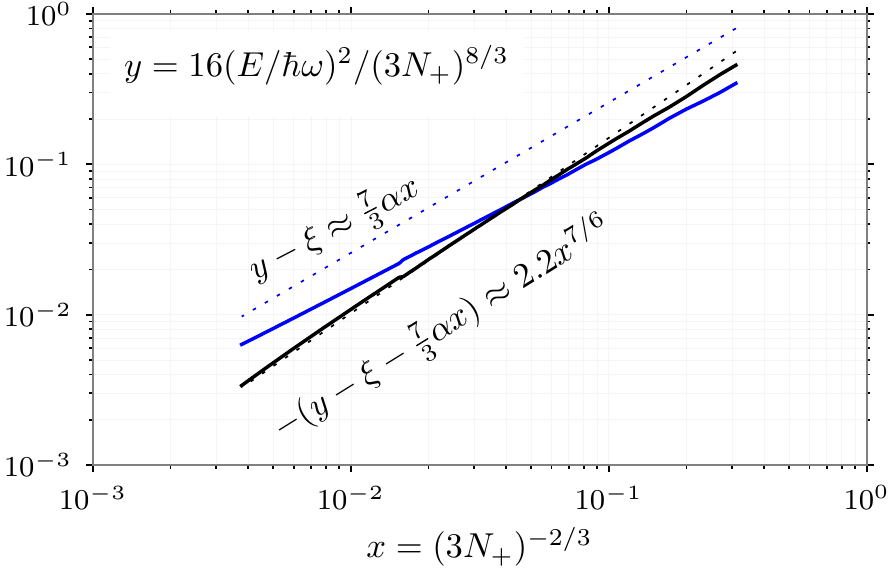}
  \caption{(color online) 
    Asymptotic behaviour of the \gls{SLDA} model with $\alpha = \num{1.104(8)}$,
    $\xi = \num{0.3742(5)}$, and $\eta = \num{0.651(1)}$: $y = \xi + 7\alpha x/3
    + a x^{7/6} + \cdots$ where $a\approx -2.2$.  We plot the deviations $y-\xi$
    (upper solid blue curve) and $-(y - \xi - 7\alpha x/3)$ (lower solid black
    curve) along with the missing correction terms $7\alpha x/3$ (upper dotted
    blue line) and $2.2 x^{7/6}$ (lower dotted black line) to demonstrate the
    asymptotic powers.  Note that the scale is quite different from
    figure~\ref{fig:ETF_asympt} owing to the additional computational complexity
    of the \gls{SLDA}.  The approach to the asymptotic form is consistent.}
  \label{fig:SLDA_asympt} 
\end{figure}

\noindent
Figure \ref{fig:ETF_asympt} is a full-sized version of the inset of
Fig.~\ref{fig:ETF_SLDA}.  This shows on a log-log scale the deviation of the
numerically computed $y(x)$ from the \gls{ETF} model from the expected
asymptotic form~\eqref{eq:Etrap_x} $y= \xi + \cSR x - 2.2x^{7/6} + \cdots$. (We
do not claim anything universal about the coefficient $2.2$ here.  It was simply
obtained numerically by fitting the asymptotic form: likely, a more careful
analysis of the \gls{SLDA} results will yield a different value.)  This confirms
that the \gls{ETF} does approach the asymptotic form, but only for $N_+ \sim
10^8$ as expected from the proximity of the correction.  We also show the same
analysis for the \gls{SLDA} in Fig.~\ref{fig:SLDA_asympt}, however, since it is
much more expensive numerically, we have not approached the $N_+\approx 10^8$
threshold (note the different scales).  Although the agreement does not seem as
good, this is simply a function of the small particle numbers: a comparison with
the \gls{ETF} results in the same region shows similar deviations.

\section{No We\"\i{}sacker Term in the ETF Model}
\label{sec:hydr-with-etf}
\noindent
Superfluid hydrodynamic simulations with the \gls{ETF} find phenomenologically
that the \gls{ETF} without gradient corrections works best to describe
collisional dynamics in the \gls{UFG}~\cite{Ancilotto:2012, *Ancilotto:2012a}.
Here we also argue that they must vanish to give a sensible description of
vortices. The \gls{ETF} model~\eqref{eq:ETF} is equivalent to a modified
\gls{GPE}~\cite{Salasnich:2010, Forbes:2012b} with a complex field $\Psi$
describing dimers of mass $2m$ normalized such that the density $n_+ =
2\abs{\Psi}^2$:
\begin{gather*}
  \I\partial_t\Psi = \left[\frac{-\hbar^2\nabla^2}{4m} + 2(\xi E_F(n_+) - \mu)
    - \dlambda\frac{\hbar^2\nabla^2\abs{\Psi}}{m\abs{\Psi}}\right]\Psi.
\end{gather*}
Consider a single stationary vortex $\Psi \propto e^{\I\phi}f(r)$ embedded in a
uniform gas with background chemical potential $\mu = \xi E_F(n_+^{\infty}) = b
f^{4/3}(r=\infty)$.  The phase yields a centrifugal term:
\begin{gather}\label{eq:centrifugal}
  \left(
    \frac{\hbar^2}{4m r^2}
    - \frac{(1 + 4\dlambda)\hbar^2\nabla^2}{4m}
    + bf^{4/3}(r) - \mu
  \right) f(r) = 0.
\end{gather}
The Weizsäcker term modifies the effective mass in the gradient term, but does
not similarly alter the centrifugal term (the first term in
Eq.~\eqref{eq:centrifugal}) since it acts only on the modulus $\abs{\Psi}$.  The
vortex develops a cusp near the core: let $f(r) = a r^{\alpha}$ where $\alpha
\approx 1$; then
\begin{gather*}
  \frac{ar^{\alpha-2}}{4m}-
  \frac{1 + 4\dlambda}{4m} a\alpha^2r^{\alpha-2}
  = 
  b a^{7/3} r^{7\alpha/3} - \mu a r^{\alpha}
\end{gather*}
The terms on the left-hand-side are divergent and must cancel, fixing
$\alpha^{-2} = 1 + 2\dlambda$.  With no Weizsäcker term $\dlambda=0$, this
yields the familiar $\alpha = 1$, but the presence of a Weizsäcker spoils the
cancellation between the gradient and centrifugal terms, and the density profile
of a vortex at the core becomes non-analytic with a cusp of a fractional
power. This non-analytic cusp causes unphysical dynamical evolution of the
vortex.

\section{Naïve Gradient Corrections}\label{sec:naive-grad-corr}
\noindent
To show that a negative gradient correction is somewhat surprising, consider
expanding a local interaction in terms of the separation $\vr = (\vx - \vy)/2$
in the spirit of the \gls{DME}~\cite{Negele:1972,*Negele:1975}:  
\begin{subequations}
  \begin{gather*}
    V(\vx - \vy)n(\vx)n(\vy) \sim  V(2\vr)n(\vR + \vr)n(\vR - \vr)
    \\
    \sim V(2\vr)n^2(\vR) - 2V(2\vr)\bigl[\vr \cdot \vnabla n(\vR)\bigr]^2
  \end{gather*}
\end{subequations}
Thus, a naïve expansion for an attractive potential would imply a positive
gradient correction.  It is apparent that this simplistic argument does not
apply to the strongly interacting \gls{UFG}.

\end{document}